\documentclass[aps,pra,onecolumn,groupedaddress,showpacs,showkeys]{revtex4}

\usepackage{graphicx}

\begin{document}

\newcommand{\be}{\begin{equation}}
\newcommand{\ee}{\end{equation}}

\title{The thermal Casimir effect for conducting plates\\
and the Bohr-van Leeuwen theorem.}

\author{Giuseppe Bimonte}

\affiliation{Dipartimento di Scienze Fisiche Universit\`{a} di
Napoli Federico II Complesso Universitario MSA, Via Cintia I-80126
Napoli Italy and INFN Sezione di Napoli, ITALY}

\begin{abstract}
We examine recent prescriptions for estimating the thermal Casimir
force between two metallic plates from the point of view of their
consistency with the Bohr-van Leeuwen theorem of classical
statistical physics. We find that prescriptions including the
effect of ohmic dissipation  satisfy the theorem, while
prescriptions that neglect ohmic dissipation do not.
\end{abstract}

\keywords{Casimir, thermal, statistical}
\maketitle

\section{Introduction}

Much attention has been devoted recently to the study of the
influence of temperature on the Casimir force \cite{Mohid}. In
this contribution, we shall focuss our attention on the thermal
Casimir force for two metallic, non-magnetic plates, a problem
that involves subtle theoretical difficulties not yet understood
as we write.
We recall that the thermal Casimir pressure $P(a,T)$ between two
plane-parallel dielectric plates in vacuum at distance $a$ is
provided by the following formula due to Lifshitz: \be P(a,T)=
-\frac{k_B T}{2 \pi^2} \sum_{l \ge
0}\left(1-\frac{1}{2}\delta_{l,0} \right) \int \!\! d^2{\bf
k_{\perp}} q_l\!\!\!\! \sum_{\alpha={\rm TE,TM}} \left(\frac{e^{2
a q_l}}{r_{\alpha}^2(i \xi_l,k_{\perp})} -1
\right)^{-1},\label{lifs} \ee  where ${\bf k_{\perp}}$ is the
on-plane wave-vector, $\xi_l= 2 \pi k_B T l/\hbar$ are the
Matsubara frequencies, $q_l =\sqrt{k_{\perp}^2+\xi_l^2/c^2}$, and
$r_{\alpha} (i \xi_l, k_{\perp})$  are the reflection coefficient
of the slabs, evaluated at imaginary frequencies $\omega_l=i
\,\xi_l$.
In the case of conducting plates, the object of controversy is the
correct magnitude of the terms with $l=0$, whose evaluation
requires making a prescription for the reflection coefficients
${r_{\alpha}(0,k_{\perp})}$ of the slabs at zero-frequency (the
terms with $l>0$ pose no problems, and can be evaluated very
accurately using optical data of the material constituting the
plates). We shall only consider in what follows normal (i.e. non
superconducting) non-magnetic good conductors, like gold (poor
conductors have also recently attracted much attention
\cite{lamor}, but we shall not consider them here). For good
conductors ${r_{\rm TM}(0,k_{\perp})}$ is obviously equal to one,
but the correct magnitude of ${r_{\rm TE}(0,k_{\perp})}$ is
controversial. The alternative prescriptions that have been
proposed for ${r_{\rm TE}(0,k_{\perp})}$ can be dubbed as the
Drude and the plasma prescription, respectively. On one hand,  the
Drude prescription maintains that \be{r_{\rm
TE}(0,k_{\perp})}\vert_{\rm Dr}=0\;, \label{Drude}\ee which
represents the limiting value for zero-frequency of the TE Fresnel
reflection coefficient, once the  Drude model of a ohmic conductor
is adopted. On the contrary, the plasma prescription neglects
altogether ohmic dissipation and takes for ${r_{\rm
TE}(0,k_{\perp})}$ the following non-vanishing value: \be r_{\rm
TE}(0,{\bf k_{\perp}})\vert_{\rm pl}= \frac{
\sqrt{\Omega_P^2/c^2+k_{\perp}^2}-k_{\perp}}{
\sqrt{\Omega_P^2/c^2+k_{\perp}^2}+k_{\perp}}\;, \label{plamod}\ee
where $\Omega_P$ is the  plasma frequency of the conductor, as
obtained from optical data at IR frequencies.  There is no room
here to discuss in detail the reasons in favor or against either
prescription, and we address the reader to Ref.\cite{Mohid}.

In order to discriminate between the Drude and the plasma
prescriptions, we recently proposed \cite{bimonte} to test their
consistency with a well known result from classical statistical
physics, i.e. the Bohr- van Leeuwen theorem \cite{van}. To justify
this approach, we note on one hand that the Casimir effect is an
equilibrium phenomenon, and therefore the theoretical models used
for the plates should be consistent with the dictates of
statistical physics. It is interesting to note, in this regard,
that statistical physics does indeed imply very general
constraints, known as Onsager's reciprocity relations, on the
possible form of the $2 \times 2$ reflection matrix of a
homogeneous possibly anisotropic surface, that for example rule
out certain phenomenological models of chiral materials
\cite{santa}. On the other hand, to justify recourse to a theorem
of classical statistical physics, we observe that, differently
from the $l>0$ terms, the troublesome $l=0$ terms of Lifshitz
formula have essentially a classical character, since they do not
explicitly involve Planck's constant. Obviously this remark
applies only to normal metals,  and indeed the Bohr- van Leeuwen
theorem was originally derived to explain their weak magnetic
properties. Obviously, this is not the case for magnetic or
superconducting materials, whose response functions depend on
quantum effects that disappear in the classical limit.

\section{The Bohr-van Leeuwen theorem}

The Bohr-van Leeuwen theorem originated early in the 20th century,
in an attempt to explain the absence of strong diamagnetism in
normal conductors placed in an external magnetic field \cite{van}.
By slightly generalizing its initial content, that referred to a
material placed in a static external magnetic field, we can state
the theorem as saying that in classical systems at thermal
equilibrium {\it matter decouples from transverse em fields}. To
prove it, consider the {\it classical microscopic} Hamiltonian for
a system of $N$ charged particles interacting with the em field,
in the Coulomb gauge (${\bf \nabla}\cdot {\bf A}=0$): $$
H=\sum_{i=1}^N \frac{1}{2 m_i}\left[{\bf P}_i-\frac{e_i}{c}({\bf
A}({\bf r}_i)+{\bf A}_{\rm ext}({\bf r}_i))\right]^2+\sum_{i
<j}e_i e_j v(|{\bf r}_i-{\bf r}_j|)$$ \be+\sum_{i=1}^N V^{({\rm
walls})}({\bf r}_i)+H_{0}^{({\rm rad})}\;,\ee where $v(|{\bf
r}_i-{\bf r}_j|)$ is the Coulomb electrostatic potential,
$V^{({\rm walls})}({\bf r})$ is a confining potential mimicking
the walls of the particles' containers, $H_{0}^{({\rm rad})}$ is
the free Hamiltonian for the fluctuating em field ${\bf A}$ and we
possibly allow for the presence of an external static magnetic
field with potential ${\bf A}_{\rm ext}$. Consider now the
classical partition function of the system $Z=\int d\mu_{\rm
part}\int d\mu_{\rm rad} \exp{(-\beta H)}$, where $d \mu_{\rm
part}$ and $d \mu_{\rm rad}$ are the phase-space canonical
measures  for the particles and the em field, respectively. It is
then straightforward to check that under the canonical
transformation: \be {\bf P}_i \rightarrow {\bf P}_i'={\bf
P}_i-\frac{e_i}{c}\left({\bf A}({\bf r}_i)+{\bf A}_{\rm ext}({\bf
r}_i)\right)\ee $Z$  {\it factorizes} into the product \be
Z=Z^{({\rm part})} \times Z_0^{(\rm rad)}\;,\label{fact}\ee where
$Z_0^{(\rm rad)}=\int d\mu_{\rm rad} \exp{(-\beta H_0^{({\rm
rad})})}$ is the partition function for the free em field in empty
space, and $Z^{(\rm part )}=\int d\mu_{\rm part} \exp{(-\beta
H^{(\rm part )})}$, where $ H^{(\rm part )}=\sum_{i=1}^N {\bf
P}_i^2/(2 m_i) +\sum_{i <j}e_i e_j v(|{\bf r}_i-{\bf r}_j|)$. The
factorization property Eq. (\ref{fact}) shows that at {\it thermal
equilibrium} the particles and the transverse em field are
completely decoupled. This decoupling property has two important
physical consequences: on one hand, it implies that an external
magnetic field  ${\bf A}_{\rm ext}$ does not affect the
thermodynamic properties of matter, and on the other hand it
implies that the presence of matter has no influence on thermal
averages of the fluctuating em field. The former implication
provides the original content of the Bohr-van Leeuwen theorem
\cite{van}, while the second implication is  the one we need for
the purposes of Casimir physics, as we discuss below.

\section{The Bohr-van Leewuen theorem and the Casimir effect}

In the previous Section, we have shown that in classical
statistical physics, the thermal averages of the transverse em
field are independent of matter.
As we now explain, this result has important consequences for the
thermal Casimir effect. We consider the simple case of two
identical plane-parallel homogeneous and isotropic slabs with
complex permittivity $\epsilon(\omega)$, lying in the $(x,y)$
plane, separated by an empty gap of width $a$. As it is well
known, the Casimir pressure $P(a,T)$ between the slabs is equal to
the (renormalized) quantum thermal average $\langle T_{zz}
\rangle$ of the component $T_{zz}$ of the Maxwell stress tensor,
evaluated at any point in the empty gap. It is possible to split
$\langle T_{zz} \rangle$ as \be \langle T_{zz} \rangle=\langle
T_{\| zz} \rangle+\langle T_{\perp zz} \rangle \;,\ee where
$\langle T_{\| zz} \rangle$ and $\langle T_{\perp zz} \rangle$
represent the contributions from the longitudinal and the
transverse em field, respectively. It is shown in Ref.
\cite{bimonte} that $\langle T_{\| zz} \rangle$ and $\langle
T_{\perp zz} \rangle$ have the expressions: \be \langle T_{\| zz}
\rangle=\frac{1}{\pi^2}\int_0^{\infty} \!\!\frac{d\omega}{\omega}
E_{\beta}(\omega)\!\int {d{k}_{\perp}} \,k_{\perp}^2 \, {\rm
Im}\left[\left(1- \frac{e^{2 k_{\perp} d}}{{\bar
r}^2(\omega)}\right)^{-1}\right],\ee \be \langle T_{\perp
zz}\rangle=-\frac{2}{\pi}\int_0^{\infty}
\!\!\frac{d\omega}{\omega} E_{\beta}(\omega)\,{\rm Im}\, [{\cal
T}_{\perp}(\omega)]\,\label{Tzztra}\ee where \be {\cal
T}_{\perp}(\omega)= \frac{1}{2 \pi}\!\int_0^{\infty}
{d{k}_{\perp}} \,k_{\perp} \left\{q \sum_{\alpha={\rm TE,
TM}}\left( \frac{e^{2 q d}}{{r}_{\alpha}^2(\omega,
k_{\perp})}-1\right)^{-1}\!\!\!- k_{\perp}\left( \frac{e^{2
k_{\perp} d}}{{\bar r}^2(\omega)}-1\right)^{-1}\right\}\;.
\label{calT}\ee Here $q=\sqrt{k_{\perp}^2-\omega^2/c^2}$,
 ${\bar r}=({\epsilon(\omega)-1})/({\epsilon(\omega)+1})$ and
$E_{\beta}(\omega)={\hbar \omega}/{2} \coth[{\hbar \omega}/{(2 k_B
T)}]$.   Consider now the transverse contribution $\langle
T_{\perp zz} \rangle$: since, as shown in the previous Section,
the classical statistical averages of the transverse em field are
independent of matter, it follows that $\langle T_{\perp zz}
\rangle$ must vanish in the classical limit. For $\hbar
\rightarrow 0$, the quantity $E_{\beta} (\omega)$ approaches the
limit $k_B T$, and then it can be taken outside the integral on
the r.h.s. of Eq. (\ref{Tzztra}). After further rotating the
$\omega$ domain of integration from the real positive axis to the
imaginary positive axis (such a rotation is permitted by the
analyticity properties of the complex permittivity
$\epsilon(\omega)$ of causal media), it is easy to verify
\cite{bimonte} that \be \lim_{\hbar \rightarrow 0} \langle
T_{\perp zz}\rangle=- \,k_B T\, \lim_{\omega \rightarrow 0}\,
{\cal T}_{\perp}(\omega)\;.\label{Tperpzzclbis}\ee It can be
easily verified that for vanishing frequency the ${\rm TM}$
contribution to ${\cal T}_{\perp}(\omega)$ always cancels against
the second term between the curly brackets on the r.h.s. of Eq.
(\ref{calT}), and then one finds: \be \lim_{\hbar \rightarrow 0}
\langle T_{\perp zz}\rangle=- \frac{\,k_B T}{2
\pi}\,\int_0^{\infty} {d{k}_{\perp}} \,k^2_{\perp} \left(
\frac{e^{2 k_{\perp} d}}{{r}_{\rm TE}^2(0,k_{\perp})}
 -1\right)^{-1}\;.\label{TEclas}\ee
The Bohr-van Leeuwen  theorem requires that the quantity on the
r.h.s. vanishes for all separations, and this is only possible if
\be {r}_{\rm TE}(0,k_{\perp})=0\;.\ee The conclusion is that the
Drude prescription Eq. (\ref{Drude}) is consistent with the
theorem, while the plasma prescription Eq. (\ref{plamod}) is not.

\section{Concluding remarks}

The Bohr-van Leeuwen theorem of classical statistical physics
states that in the classical limit the transverse em field
decouples from matter. We have used this theorem as a criterion to
discriminate between the so-called Drude and plasma prescriptions
that have been recently proposed to evaluate the thermal Casimir
force between two metallic non-magnetic plates. We have shown that
the Drude prescription is consistent with this theorem, while the
plasma prescription is not. The results derived in this paper do
not apply to magnetic or superconducting materials, because the
properties of such materials arise from quantum effects that
disappear in the classical limit.


\begin{thebibliography}{99}

\bibitem{Mohid} G.L. Klimchitskaya, U. Mohideen and V.M.
Mostepanenko,
arXiv:0902.4022.



\bibitem{lamor} L.P. Pitaevskii, Phys. Rev. Lett.
{\bf 101}, 163202 (2008).


\bibitem{bimonte} G. Bimonte, Phys. Rev. A {\bf 79}, 042107 (2009).

\bibitem{van} H. J., van Leeuwen, J. Phys. Radium {\bf 2}, 362 (1921);
J. H. Van Vleck, {\it The Theory of Electric and Magnetic
Susceptibilities} (Clarendon Press, Oxford, 1932).

\bibitem{santa} G. Bimonte and E. Santamato, Phys. Rev. A {\bf 76},
013810 (2007).





\end{thebibliography}
\end{document}